\documentclass[twocolumn,aps,prd,preprintnumbers,showpacs,superscriptaddress,nofootinbib,amsmath,amssymb,floats,floatfix,showkeys,notitlepage,longbibliography]{revtex4}

\usepackage[]{mdframed}

\usepackage{braket,amssymb,amsmath,array,commath,mathtext,wrapfig,tikz}

\usepackage{bm}
\usepackage{amsthm,amsfonts,amscd}
\usepackage{graphicx}
\usepackage{lscape}

\usepackage{hyperref}
\hypersetup{colorlinks=true,linkcolor=blue,urlcolor=blue,citecolor=blue}

\begin{document}

\title{Analyzing the Influence of Geometrical Deformation on Photon Sphere and Shadow Radius: A New Analytical Approach - Spherically Symmetric Spacetimes}

\author{Vitalii Vertogradov}
\email{vdvertogradov@gmail.com}

\affiliation{Physics department, Herzen state Pedagogical University of Russia,
48 Moika Emb., Saint Petersburg 191186, Russia} 
\affiliation{SPB branch of SAO RAS, 65 Pulkovskoe Rd, Saint Petersburg
196140, Russia}

\author{Ali \"Ovg\"un
}
\email{ali.ovgun@emu.edu.tr}
\affiliation{Physics Department, Eastern Mediterranean
University, Famagusta, 99628 North Cyprus, via Mersin 10, Turkiye}

\begin{abstract}
In this paper, we introduce a new approach to study the behavior of the photon sphere and shadow radius. Our method uses extended gravitational decoupling and reveals two important analytic results. First,  the additional matter field alters the photon sphere radius: it increases if $g'(r_{ph}^{(0)})>0$ and decreases if $g'(r_{ph}^{(0)})<0$ (where $g'$ represents the derivative of a specific metric function evaluated at the original photon sphere radius). Second, the presence of the matter field can modify the black hole shadow size. If $g\left(r_{ph}^{(0)}\right)>0$, the shadow shrinks, while it grows for $g\left(r_{ph}^{(0)}\right)<0$. These findings provide a deeper insight into how matter distribution influences the characteristics of black holes and their observable features. Through a systematic framework and various illustrative examples, our investigation not only clarifies these fundamental aspects but also significantly enhances the theoretical framework of black hole astrophysics.

\end{abstract}

\date{\today}



\maketitle

\renewcommand{\refname}{References}

\section{Introduction}
One of the most convincing pieces of experimental evidence for the existence of black holes comes from the images of a black hole's shadow, captured by the Event Horizon Telescope Collaboration~\cite{bib:eht1, bib:eht2, bib:eht3}. The pioneering work on light deflection by gravitationally intense stars is attributed to Synge~\cite{bib:singe}. Bardeen calculated the radius of a photon sphere for a Schwarzschild black hole, demonstrating that its radius equals $3M$~\cite{bib:bardeen1}. This photon sphere manifests as a shadow, appearing as a dark spot in the observer's sky, with a radius of $3\sqrt{3}M$ (roughly 2.5 times larger than the radius of the event horizon, $r=2M$). He also demonstrated that in the case of a rotating object, the shape of the shadow undergoes changes. However, in reality, astrophysical black holes are not isolated objects; they exist within an expanding universe and are often surrounded by luminous accretion flows, factors that significantly influence our observations. Therefore, to accurately describe a black hole shadow, we must take into account a black hole surrounded by an accretion disk. Perlick and Tsupko et al. extensively investigated the impact of plasma on a black hole shadow~\cite{bib:tsupko_plasma,bib:tsupko_deflection,bib:tsupko_lensing,bib:tsupko_kerr} . For further insights, readers may refer to a comprehensive review on black hole shadows by Tsupko~\cite{bib:tsupko_review}, which also includes relevant references. \textcolor{black}{The idea of using millimeter-wavelength global interferometers to image the black hole shadow at the Galactic Center was first proposed in \cite{ZAKHAROV2005479}. This prediction turned J. Bardeen's thought experiment into a realized test of General Relativity \cite{Zakharov:2023yjl}. An analytical expression for the shadow size as a function of charge was derived in \cite{Zakharov:2005ek}, with the corresponding (blue) curve presented in \cite{EventHorizonTelescope:2021dqv}. Additionally, the authors used this relationship to explore the existence of a black hole in higher-dimensional space \cite{2012NewAR..56...64Z}. Recent studies on M87* and Sgr A* utilized these analytical expressions for the shadow in the Reissner-Nordstrom metric to constrain the tidal charges of these black holes \cite{Zakharov:2022gwk, Zakharov:2021gbg}. While it is known that photon spheres influence black hole shadows \cite{Zakharov:1986a}, a formal demonstration is needed to confirm that, for the Schwarzschild, Kerr, and Reissner-Nordstrom metrics, photons with impact parameters smaller than the critical value (corresponding to the photon sphere) are indeed captured by the black hole \cite{Zakharov:1994}.} The diverse characteristics of static, stationary, and dynamical black holes have been thoroughly investigated in a series of studies~\cite{Bambhaniya:2022xbz,bib:klaudel,bib:hod,Virbhadra:2024pru,Adler:2022qtb,Virbhadra:2022iiy,Virbhadra:2007kw,Virbhadra:1999nm,Virbhadra:1998dy, bib:khoo, bib:decaniny, bib:shoom, bib:sederbaum, bib:jokhannsen, bib:teo,Okyay:2021nnh,Pantig:2022gih,Kuang:2022xjp,Ovgun:2024zmt, Karshiboev:2024xxx, Uniyal:2023ahv, Pulice:2023dqw, Yang:2023tip, Cimdiker:2023zdi, Kumaran:2023brp, Lambiase:2023hng, Ovgun:2023ego, Atamurotov:2022knb, Kumaran:2022soh, Pantig:2022qak, Cimdiker:2021cpz, Ovgun:2020gjz, Ovgun:2018tua,Zakharov:2023lib,Qiao:2022hfv,Hod:2020pim,Lu:2019zxb,Paithankar:2023ofw,Aratore:2024bro,Tsupko:2022kwi,bib:rezzolla, bib:rezzolla2, bib:expanding, bib:vaidya1, bib:vaidya2, bib:ver_shadow, bib:tsupko_first, bib:understanding, bib:joshi_shadow,Tsukamoto:2014tja,Tsukamoto:2017fxq,Gomez:2024ack,Adair:2020vso,Ghosh:2023kge,Pedrotti:2024znu,Afrin:2022ztr,Chen:2022nbb}. The black hole shadow can serve as a cosmological ruler~\cite{bib:tsupko_ruller, bib:vanozi_ruller} and as a means to test various black hole models~\cite{bib:vanozi}.

Indeed, one approach involves considering a black hole surrounded by plasma and examining how light propagation within it affects the angular size of the shadow. Alternatively, the black hole-accretion disk system can be treated as a solution of the Einstein field equations, allowing for the analysis of light trajectories within the obtained spacetime.
For instance, investigations into the shadow and optical appearance of a black hole surrounded by Chaplygin gas have been conducted~\cite{bib:chaplygin}. It is expected that an accretion disk would have minimal influence on the exterior geometry of a black hole, with any small deformations potentially being described by minimal geometrical deformation~\cite{bib:mgd1, bib:mgd2} or extended gravitational decoupling~\cite{bib:gd1, bib:gd2, bib:gd3}. If we exclusively consider deformations in the $g_{11}$ component of the metric tensor, which corresponds to minimal geometrical deformations, they would not affect the radius of the photon sphere or the visible size of the shadow. Therefore, it becomes necessary to account for deformations in both the $g_{00}$ and $g_{11}$ components of the metric tensor, a scenario well-described by extended gravitational decoupling. Using gravitational decoupling, several new solutions of the Einstein field equations have been discovered, depicting deformed versions of well-known black hole solutions~\cite{bib:bh1, bib:bh2, bib:ovalle_regular, bib:max, bib:max_regular, bib:kiselev, bib:rotating, bib:varmhole}. These deformations arise due to the presence of an additional anisotropic matter field surrounding the black hole.


Understanding the impact of an additional anisotropic matter field on black hole solutions is crucial for refining our comprehension of these cosmic entities. Such deformations to general spherically-symmetric static black holes have far-reaching consequences, potentially altering fundamental characteristics such as the radius of the photon sphere, the structure of the black hole shadow, and its angular size. By delving into this phenomenon, we aim to shed light on the intricate interplay between matter distribution and gravitational effects around black holes. Through a comprehensive examination supported by various illustrative examples, our study contributes to advancing the theoretical framework underlying black hole astrophysics.

This paper is structured as follows: In Section II, we provide a concise overview of extended gravitational decoupling. Section III is dedicated to estimating the impact of deformations induced by an additional anisotropic matter field on the radius of the photon sphere and the resulting shadow. The discussion of our findings is presented in Section IV.

Throughout the paper, we adopt the system of geometrized units with $G=c=1$, and employ the signature $-+++$ consistently.

\section{Gravitational decoupling}

In this section, we provide a brief overview of gravitational decoupling. The concept of extended gravitational decoupling posits that, under specific conditions, it is feasible to resolve the Einstein field equations with two distinct matter sources.
\begin{equation}
\tilde{T}_{ik}=T_{ik}+\Theta_{ik} \,,
\end{equation}
where $T_{ik}$ represents the energy-momentum tensor of a system for which the Einstein field equations are applied.
\begin{equation} \label{eq:thefirst}
G_{ik}=8\pi T_{ik} \,.
\end{equation}
The solution to \eqref{eq:thefirst} is assumed to be familiar, referred to as a seed solution. This seed solution may correspond to well-known solutions of the Einstein field equations, such as Schwarzschild, Reissner-Nordström, Hayward, Bardeen, among others. $\Theta_{ik}$ denotes an additional matter source responsible for further geometrical distortions. This matter source adheres to the Einstein field equations, defined by:
\begin{equation} \label{eq:thesecond}
\bar{G}_{ik}=\alpha \Theta_{ik} \,,
\end{equation}
where $\alpha$ represents a coupling constant, and $\bar{G}_{ik}$ denotes the Einstein tensor of the deformed metric.

The Einstein field equations are inherently nonlinear differential equations. However, gravitational decoupling enables the solution of the Einstein field equations \eqref{eq:thefirst} and \eqref{eq:thesecond} under certain conditions. Consequently, a direct superposition of these two solutions is possible
\begin{equation}
\tilde{G}_{ik}\equiv G_{ik}+\bar{G}_{ik}=8\pi T_{ik}+\alpha
\Theta_{ik}\equiv \tilde{T}_{ik} \,,
\end{equation}
is also a valid solution of the Einstein field equations.

We consider the following field equations:
\begin{equation} \label{eq:ex1}
G_{ik}=R_{ik}-\frac{1}{2}g_{ik}R=8\pi T_{ik} \,.
\end{equation}
Let the solution of \eqref{eq:ex1} represent a static spherically-symmetric spacetime of the form
\begin{equation} \label{eq:seed}
ds^2=-e^{\nu(r)}dt^2+e^{\lambda(r)}dr^2+r^2 d\Omega^2 \,.
\end{equation}
Here, $d\Omega^2=d\theta^2+\sin^2\theta d\varphi^2$ denotes the metric on the unit two-sphere. The functions $\nu(r)$ and $\lambda(r)$ are dependent on the radial coordinate $r$ and are assumed to be known.
The metric given in \eqref{eq:seed} is referred to as the seed metric.

Now, we aim to investigate the geometrical deformation of \eqref{eq:seed} by introducing two new functions $g=g(r)$ and $f=f(r)$, defined as:

\begin{eqnarray} \label{eq:deform}
e^{\nu(r)} &\rightarrow & e^{\nu(r)+\alpha g(r)},\nonumber \\
e^{\lambda(r)} &\rightarrow & e^{\lambda(r)}+\alpha f(r).
\end{eqnarray}

Here, $\alpha$ represents a coupling constant. Functions $g$ and $f$ are linked to the geometrical deformations of $g_{00}$ and $g_{11}$ of the metric \eqref{eq:seed}, respectively. These deformations are induced by the new matter source $\Theta_{ik}$. If we set $g(r)\equiv 0$, then only the $g_{11}$ component undergoes deformation, while $g_{00}$ remains unchanged. This represents the minimal geometrical deformation. However, this approach has certain limitations. For instance, it becomes challenging to achieve a stable black hole configuration with a well-defined event horizon using this method (see Ref.~\cite{bib:bh2} for details). If we deform both the $g_{00}$ and $g_{11}$ components, this constitutes an extended gravitational decoupling approach.

Substituting \eqref{eq:deform} into \eqref{eq:seed}, we obtain:
\begin{equation} \label{eq:noseed}
ds^2=-e^{\nu+\alpha g}dt^2+\left(e^{\lambda}+\alpha f\right)
dr^2+r^2 d\Omega^2 \,.
\end{equation}

The Einstein equations for \eqref{eq:noseed}
\begin{equation}
\tilde{G}_{ik}= 8 \pi \tilde{T}_{ik}=8\pi (T_{ik}+\Theta_{ik} ) \,,
\end{equation}
are
\begin{widetext}
\begin{eqnarray} \label{eq:einstein}
8\pi (T^0_0&+&\Theta^0_0)=-\frac{1}{r^2}+e^{-\beta}\left( \frac{1}{r^2}-\frac{\beta'}{r}\right),\nonumber \\
8\pi (T^1_1&+&\Theta^1_1)=-\frac{1}{r^2}+e^{-\beta}\left(\frac{1}{r^2}+\frac{\nu'+\alpha g'}{r}\right),\nonumber \\
8\pi (T^2_2&+&\Theta^2_2)=\frac{1}{4}e^{-\beta}\left [2(\nu''+\alpha g'')+(\nu'+\alpha g')^2-\beta'(\nu'+\alpha g')+2\frac{\nu'+\alpha g'-\beta'}{r} \right],\nonumber \\
e^{\beta}&\equiv &e^{\lambda}+\alpha f.
\end{eqnarray}
Here, the prime denotes the partial derivative with respect to the radial coordinate $r$, and $8\pi T^2_2+\Theta^2_2=8 \pi T^3_3+\Theta^3_3$ due to spherical symmetry.

\end{widetext}

From \eqref{eq:einstein}, one can define the effective energy density $\tilde{\rho}$, effective radial pressure $\tilde{P}_r$, and effective tangential pressure $\tilde{P}_t$ as:
\begin{eqnarray} \label{eq:effective}
\tilde{\rho}&=&-(T^0_0+\Theta^0_0),\nonumber \\
\tilde{P}_r&=&T^1_1+\Theta^1_1,\nonumber \\
\tilde{P}_t&=&T^2_2+\Theta^2_2 .
\end{eqnarray}

From \eqref{eq:effective}, one can introduce the anisotropy parameter $\Pi$ as:
\begin{equation} \label{eq:anisotropy}
\Pi=\tilde{P}_t-\tilde{P}_r \,.
\end{equation}

If $\Pi \neq 0$, it indicates the anisotropic behavior of the fluid $\tilde{T}_{ik}$.

\section{The influence of deformation on black hole shadow}

In this section, we consider the case when $\Lambda (r)=-\nu(r)$. Without loss of generality, we can assume

\begin{widetext}
\begin{eqnarray} \label{eq:metric}
ds^2=-\left(1-\frac{X(r)}{r}\right)e^{\alpha g(r)}dt^2+\left(1-\frac{X(r)}{r}+\alpha f(r)\right)^{-1}dr^2+r^2d\Omega^2,
\end{eqnarray}

where $X(r)$ can be interpreted as a shape function.
\end{widetext}

The spacetime \eqref{eq:metric} possesses two Killing vectors related to time-translation invariance, denoted by $k=\partial_t$, and spacelike, denoted by $n=\partial_\varphi$. If $u^i\equiv \frac{dx^i}{d\lambda}$ represents the photon four-momentum, then we have two constants of motion: the energy per mass $E$

\begin{equation} \label{eq:energy_hair}
E=-k^iu_i=\left(1-\frac{X(r)}{r}\right)e^{\alpha g(r)}\frac{dt}{d\lambda},
\end{equation}

and angular momentum per mass
\begin{equation} \label{eq:angular_hair}
L=n^iu_i=r^2\sin^2\theta \frac{d\varphi}{d\lambda}.
\end{equation}

In a spherically-symmetric spacetime, without loss of generality, one can confine the analysis to the equatorial plane $\theta=\frac{\pi}{2}$. From the condition for lightlike geodesics, $g_{ik}u^iu^k=0$, one can derive the radial component of the four-velocity

\begin{equation} \label{eq:radial}
\frac{\left(1-\frac{X(r)}{r}\right)e^{\alpha g(r)}}{1-\frac{X(r)}{r}+\alpha f(r)}\left(\frac{dr}{d\lambda}\right)^2+V_{eff}(r, L, E)=0,
\end{equation}
where
\begin{equation} \label{eq:potential}
V_{eff}(r, L, E)=\left(1-\frac{X(r)}{r}\right)e^{\alpha g(r)}\frac{L^2}{r^2}-E^2,
\end{equation}
is the effective potential. 

We are interested in a circular light orbit, which implies $\frac{dr}{d\lambda}=0$ and $\frac{d^2r}{d\lambda^2}=0$. This leads us to the following conditions on the effective potential:

\begin{equation} \label{eq:conditions}
V_{eff}(r_{ph})=0,~~ V'_{eff}(r_{ph})=0,
\end{equation}

where $r_{ph}$ is the radius of a photon sphere, and the dash denotes a derivative with respect to $r$. Solving the second equation \eqref{eq:conditions} with respect to radius and substituting it into the first condition \eqref{eq:conditions}, one finds the impact parameter $B=\frac{L}{E}$. The corresponding critical impact parameter at the photon sphere is

\begin{equation}
B=\frac{r_{ph}}{\sqrt{\left(1-\frac{X(r_{ph})}{r_{ph}}\right)e^{\alpha g(r_{ph})}}}.
\end{equation}

To understand how the deformation $\alpha g(r)$ influences the radius of the photon sphere and the shadow itself, we will find these values when $\alpha = 0$. In this case, the effective potential is given by:

\begin{equation} \label{potential_nohair}
V_{eff}=\left(1-\frac{X(r)}{r}\right)\frac{L^2}{r^2}-E^2.
\end{equation}

The second condition \eqref{eq:conditions} yields the photon sphere equation:\cite{bib:klaudel} $ 
g_{00} \partial_{r} g_{\theta \theta}=g_{\theta \theta} \partial_{r} g_{00}$
leads to 
\begin{equation} \label{eq:second_nohair}
3X-X'r-2r=0,
\end{equation}

If $r_{ph}$ is the solution of the equation \eqref{eq:second_nohair}, then the first condition \eqref{eq:conditions} yields the critical impact parameter at the photon sphere

\begin{equation} \label{eq:impact_nohair}
B=\frac{r_{ph}}{\sqrt{1-\frac{X(r_{ph})}{r_{ph}}}}.
\end{equation}

The angular size of a shadow, observable by an observer at $r_o$, is given by:
\begin{equation} \label{eq:angular_nohair}
\sin^2 \alpha_{sh}=\left(1-\frac{X(r_o)}{r_o}\right)\frac{B}{r_o}.
\end{equation}

If an observer is situated far from a black hole, i.e., $r_o \gg r_{ph}$, and the spacetime is asymptotically flat, i.e., $\lim\limits_{r\to \infty} \frac{X(r)}{r}=0$, then \eqref{eq:angular_nohair} provides an approximate value for the angular size of a shadow as:
\begin{equation} \label{eq:angular_aproximate}
\alpha_{sh}\approx \frac{B}{r_o}.
\end{equation}

Now, we will consider the general effective potential \eqref{eq:potential} and explore the impact of deformation on the radius of a photon sphere, shadow, and angular size. The second condition \eqref{eq:conditions} for the effective potential \eqref{eq:potential} yields:

\begin{equation} \label{eq:second}
3X-X'r-2r+(r-X)\alpha g' r=0.
\end{equation}

If $r_{ph}$ is the solution of this equation, then the first condition \eqref{eq:conditions} yields:

\begin{equation} \label{eq:impact}
B=\frac{r_{ph}}{\sqrt{1-\frac{X(r_{ph})}{r_{ph}}}}e^{-\frac{1}{2 \alpha g(r_{ph})}}.
\end{equation}

As an example, we consider the Reissner-Nordström black hole solution, which can be obtained through gravitational decoupling \cite{bib:gd2}. The seed spacetime is assumed to be Schwarzschild, and

\begin{equation}
\alpha g(r)=\ln |1+\frac{Q^2}{r^2-2Mr}|.
\end{equation}
where $X(r)=2M$. The equation \eqref{eq:second} becomes

\begin{eqnarray}
3M&-&r+\frac{Q^2(M-r)}{r^2-2Mr+Q^2}=0, \nonumber \\
r^2&-&3Mr+2Q^2=0,\nonumber \\
r^+_{ph}&=&\frac{1}{2}\left(3M+\sqrt{9M^2-8Q^2}\right).
\end{eqnarray}

If we substitute this radius into \eqref{eq:impact}, we obtain:
\begin{equation}
B=\frac{r_{ph}^+}{\sqrt{Mr^+_{ph}-Q^2}}.
\end{equation}

However, it's important to note that in general, determining the influence of primary hair on the shadow of a black hole from equations \eqref{eq:second} and \eqref{eq:impact} can be challenging. In many cases, equation \eqref{eq:second} becomes too complex, making it impossible to find an analytical solution.

To estimate this influence, we assume that $\alpha \ll 1$. We will seek the radius of a photon sphere $r_{ph}$ as:

\begin{equation} \label{eq:radius}
r_{ph}=r_{ph}^{(0)}+\alpha r_{ph}^{(1)},
\end{equation}

where $r_{ph}^{(0)}$ is the solution of \eqref{eq:second_nohair}, i.e., the radius of the photon sphere of a seed metric.

\begin{widetext}

Substituting \eqref{eq:radius} into the second condition \eqref{eq:second} and neglecting terms of order $O(\alpha^2)$ and higher, we obtain:

\begin{equation} \label{eq:second_approx}
\left[2X'\left(r_{ph}^{(0)}\right)-X''\left(r_{ph}^{(0)}\right)r_{ph}^{(0)}-2\right]r_{ph}^{(1)}+[r_{ph}^{(0)}-X\left(r_{ph}^{(0)}\right)]g'\left(r_{ph}^{(0)}\right)r_{ph}^{(0)}=0,
\end{equation}
and one obtains for $r_{ph}^{(1)}$
\begin{equation} \label{eq:radius1}
r_{ph}^{(1)}=-\frac{[r_{ph}^{(0)}-b\left(r_{ph}^{(0)}\right)]g'\left(r_{ph}^{(0)}\right)r_{ph}^{(0)}}{2X'\left(r_{ph}^{(0)}\right)-X''\left(r_{ph}^{(0)}\right)r_{ph}^{(0)}-2}.
\end{equation}

\end{widetext}

\fbox{\begin{minipage}{25em}\textbf{ Theorem 1:} If the seed metric is the Schwarzschild one, then the additional matter field increases the radius of a photon sphere if $g'(r_{ph}^{(0)})>0$ and decreases it if $g'(r_{ph}^{(0)})<0$. \end{minipage}}
\\ \\
We observe the following cases:
\begin{itemize}
\item As an example, one can consider Reissner-Nordstrom spacetime. In this case, $r_{ph}^{(0)}=3M$ and 
\begin{equation}
\alpha g'(3M)=\frac{-4Q^2}{9M^3+3MQ^2}<0.
\end{equation}

It is a well-established fact that charge always decreases the radius of a photon sphere.

\item If the seed metric is not Schwarzschild, then we can rewrite \eqref{eq:radius1} as:

\begin{equation} \label{eq:radius_press}
r_{ph}^{(1)}=-\frac{[r_{ph}^{(0)}-X\left(r_{ph}^{(0)}\right)]g'\left(r_{ph}^{(0)}\right)r_{ph}^{(0)}}{\left(2 r_{ph}^{(0)}\right)^2\left(\rho_0+P_0\right)-2},
\end{equation}

where $\rho_0$ and $P_0$ are the energy density and pressure of the seed metric evaluated at $r=r_{ph}^{(0)}$, respectively.
The value $\left(r_{ph}^{(0)}\right)^2\rho_0$ should be less than $1$ outside the event horizon~\cite{bib:wisser_dirty}. Moreover, to satisfy the dominant energy condition, $|P_0|\leq \rho_0$. Thus, one can write:

\begin{equation}
4X'-2=\frac{12X''}{r_{ph}^{(0)}}-10\leq 0.
\end{equation}

and, as in the Schwarzschild case, the additional matter field increases the radius of a photon sphere if: $g'\left(r_{ph}^{(0)}\right)>0$ and decreases it if $g'\left(r_{ph}^{(0)}\right)<0$. 
\end{itemize}

In order to estimate the influence of geometrical deformation $\alpha g$ on an impact parameter $B$ \eqref{eq:impact}, and consequently, on the angular size of a shadow \eqref{eq:angular_nohair} and \eqref{eq:angular_aproximate}, we express the impact parameter as:

\begin{equation} \label{eq:prom}
B=B_0+\alpha \frac{\partial B}{\partial \alpha}|_{\alpha=0}+O(\alpha^2),
\end{equation}

where $B_0$ is the impact parameter of the seed metric, and:

\begin{equation} 
\frac{\partial B}{\partial \alpha}|_{\alpha=0}=-\frac{B_0 \alpha g\left(r_{ph}^{(0)}\right)}{2\left[1-\frac{X\left(r_{ph}^{(0)}\right)}{r_{ph}^{(0)}}\right]}.
\end{equation}

If we substitute this expression into (\ref{eq:prom}), one obtains
\begin{eqnarray} \label{eq:tsupko}
B&=&B_0\left[1-\alpha g\left(r_{ph}^{(0)}\right)  \beta_0\right], \nonumber \\
\beta_0&=&\frac{1}{2}\frac{1}{1-\frac{X\left(r_{ph}^{(0)}\right)}{r_{ph}^{(0)}}}>0.
\end{eqnarray}

\fbox{\begin{minipage}{25em}\textbf{ Theorem 2:} We can observe that if $g\left(r_{ph}^{(0)}\right)>0$ , then the shadow size decreases due to the additional matter field. Conversely, if 
 $g\left(r_{ph}^{(0)}\right)<0$, then the shadow increases. \end{minipage}}
\\ \\

One should note that the formula \eqref{eq:tsupko} resembles formula (17) in~\cite{bib:tsupko_plasma} with:
\begin{equation}
\alpha g\left(r_{ph}^{(0)}\right)\beta_0=\left(1-\frac{X}{r}\right)\frac{\omega_p^2}{\omega_0},
\end{equation}
where $\omega_p$ represents the plasma frequency and $\omega_0$ is the photon frequency measured by an observer at infinity.

\subsection{Example 1: Reissner-Nordstrom spacetime}

To begin, we will examine Reissner-Nordström spacetime, wherein:
\begin{equation} \label{eq:examplern}
\alpha g(r)=\ln \left[1+\frac{Q^2}{r^2-2mr}\right].
\end{equation}

In the previous analysis, it was established that for the given function $X(r)=2M$, the presence of an electrical charge $Q$ consistently decreases the radius of a photon sphere, always keeping it less than $3M$. To assess the impact of charge $Q$ on the angular size of a black hole, it's essential to evaluate $g\left( r_{ph}^{(0)}\right)$, where $r_{ph}^{(0)}=3M$. Substituting $r_{ph}^{(0)}$ into \eqref{eq:examplern} yields:

\begin{equation}
\alpha g\left(r_{ph}^{(0)}\right)=\ln \left[1+\frac{Q^2}{3M^2}\right]>0.
\end{equation}

\begin{figure}
    \centering
    \includegraphics[scale=0.8]{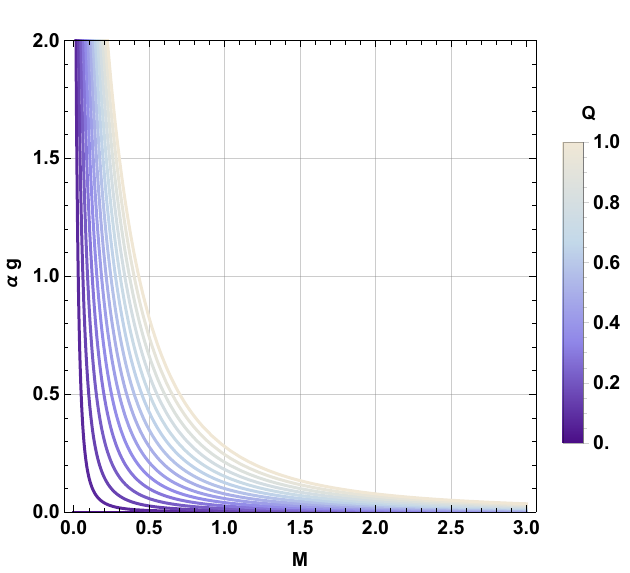}
    \caption{The Figure shows the variation of  $\alpha g\left(r_{ph}^{(0)}\right)$ due to the mass $M$ and the charge $Q$.}
    \label{fig1}
\end{figure}

Specifically, the electric charge $Q$ consistently reduces the angular size of a black hole compared to a Schwarzschild black hole. Figure \ref{fig1} illustrates the variation of $\alpha g\left(r_{ph}^{(0)}\right)$ in response to changes in both mass $M$ and charge $Q$. From Figure \ref{fig1}, it's evident that $ \alpha g\left(r_{ph}^{(0)}\right)$ remains positive throughout, indicating a decrease in the angular size of the black hole's shadow.

\subsection{Example 2: Hairy Schwarzschild black hole}

The paper~\cite{bib:bh1} presents a hairy Schwarzschild black hole solution obtained through gravitational decoupling. The solution takes the form:

\begin{eqnarray}
ds^2&=&-f(r)dt^2+f^{-1}(r)dr^2+r^2d\Omega^2,\nonumber \\
f(r)&=&1-\frac{2M+\alpha l}{r}+\alpha e^{-\frac{r}{M}}.
\end{eqnarray}

Here, $l$ is considered a primary hair with length dimension. In this case, the function $g$ is given by:
\begin{equation} \label{eq:example_hairy}
\alpha g(r)=\ln\left[1-\alpha \frac{l-re^{-\frac{r}{M}}}{r-2M}\right].
\end{equation}

In this scenario, the shape function is defined as $X(r)=2M$. Substituting $r_{ph}^{(0)}=3M$ into \eqref{eq:example_hairy}, we obtain:

\begin{equation}
\alpha g\left(r_{ph}^{(0)}\right)=\ln \left[1-\alpha\frac{l-3Me^{-3}}{M}\right].
\end{equation}

The hairy Schwarzschild spacetime satisfies the strong energy condition if the following condition is fulfilled: ~\cite{bib:bh1, bib:geod}
\begin{equation}
l-2Me^{-2}\geq 0.
\end{equation}

\begin{figure}
    \centering
    \includegraphics[scale=0.8]{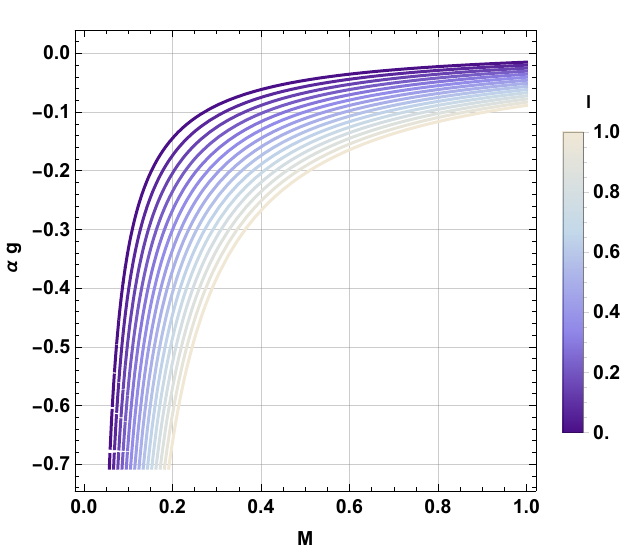}
    \caption{The figure illustrates the variation of $\alpha g\left(r_{ph}^{(0)}\right)$ with respect to the mass $M$ and the charge $l$.}
    \label{fig2}
\end{figure}

Thus, if the strong energy condition is satisfied, then $l>3Me^{-3}$, meaning the angular size of a black hole increases in this case. Figure \ref{fig2} depicts the variation of $\alpha g\left(r_{ph}^{(0)}\right)$ with respect to mass $M$ and charge $l$. As illustrated in Figure \ref{fig2}, $ \alpha g\left(r_{ph}^{(0)}\right)$ consistently remains negative, indicating an increase in the angular size of the black hole's shadow.

\subsection{Example 3: Hayward regular black hole}

As the third example, we consider a model of a regular black hole proposed by Hayward~\cite{bib:hay}. It is described by:
\begin{eqnarray}
ds^2=-f(r)dt^2+f^{-1}(r)dr^2+r^2d\Omega^2,\nonumber \\
f(r)=1-\frac{2Mr^2}{r^3+2ML^2}.
\end{eqnarray}

The regularization parameter $L$, initially assumed to be of the order of the Planck length $l_{pl}$, was later understood to be related to the non-linear electrodynamics source~\cite{bib:hay_non, bib:hay_thermo}. In this case, the function $g(r)$ is given by:
\begin{equation} \label{eq:example_hay0}
\alpha g(r)=\ln \left[1+\frac{4M^2L^2}{(r-2M)(r^3+2ML^2}\right].
\end{equation}

\begin{figure}
    \centering
    \includegraphics[scale=0.8]{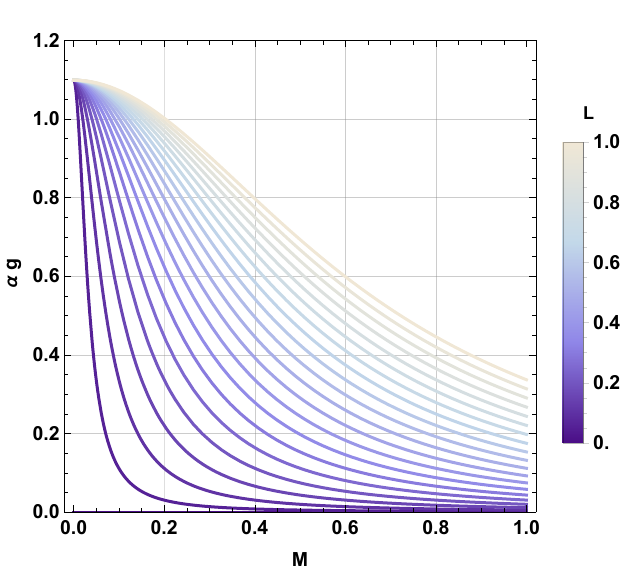}
    \caption{The Figure shows the variation of  $\alpha g\left(r_{ph}^{(0)}\right)$ due to the mass $M$ and the charge $L$.}
    \label{fig3}
\end{figure}

The seed metric is assumed to be the Schwarzschild one, i.e., $X(r)=2M$. By substituting $r_{ph}^{(0)}=3M$ into \eqref{eq:example_hay0}, we obtain:

\begin{equation}
\alpha g\left (r_{ph}^{(0)}\right)=\ln \left[1+\frac{2L^2}{4M^2+L^2}\right]>0,
\end{equation}

the parameter $L$ consistently reduces the size of a black hole shadow. Figure \ref{fig3} illustrates the variation of $\alpha g\left(r_{ph}^{(0)}\right)$ with respect to mass $M$ and charge $L$. As depicted in Figure \ref{fig3}, $ \alpha g\left(r_{ph}^{(0)}\right)$ remains positive throughout, indicating an decrease in the angular size of the black hole's shadow.

\subsection{Example 4:  Magnetically charged
Einstein-Euler-Heisenberg black hole}

The Einstein-Euler-Heisenberg nonlinear electrodynamics (NLED) arises in the low-energy limit of Born-Infeld electrodynamics. The metric function for magnetically charged black hole solutions within this theory is given by: \cite{Allahyari:2019jqz,bib:vanozi}
\begin{eqnarray}
f(r) = 1-\frac{2M}{r}+\frac{q_m^2}{r^2}-\frac{2\mu}{5}\frac{q_m^4}{r^6}\,.
\label{eq:metricmagneticallyEEH}
\end{eqnarray}
Here, $q_m$ represents the black hole's magnetic charge, which characterizes a specific hair, and $\mu$ denotes the coupling parameter in the Einstein-Euler-Heisenberg NLED, as presented in Eq.~(\ref{eq:metricmagneticallyEEH}). It's important to note that as long as $\mu \neq 0$, values of the magnetic charge $q_m>1$ are permitted.

The function $ g(r)$, in this case, is given by:
\begin{equation} \label{eq:example_4}
\alpha g(r)=\ln \left[\frac{10 M r^5+2 \mu  q^4-5 q^2 r^4-5 r^6}{5 r^5 (2 M-r)}\right].
\end{equation}

The seed metric is assumed to be the Schwarzschild one, i.e., $X(r)=2M$. Substituting $r_{ph}^{(0)}=3M$ into \eqref{eq:example_4}, we obtain:

\begin{equation}
\alpha g\left (r_{ph}^{(0)}\right)=\ln\left[1-\frac{2 \mu  q^4}{1215 M^6}+\frac{q^2}{3 M^2}\right]<0,
\end{equation}

This yields the condition $\mu > \frac{405 M^4}{2 q^2}$, meaning that if this condition is satisfied, the parameter $\mu$ decreases the angular size of a black hole shadow.

\subsection{Example 5:  Black holes in MOdified Gravity (scalar-tensor-vector gravity)}

A particularly intriguing modification of gravitational theory was proposed by Moffat in Ref.~\cite{Moffat:2005si,bib:vanozi}, commonly known as scalar-tensor-vector gravity but also referred to as MOG.

A specific black hole solution within MOG was derived in ~\cite{Moffat:2014aja}. In this scenario, the metric function describing the resulting black hole solution is given by~\cite{Moffat:2014aja}:
\begin{eqnarray}
f(r)=1-\frac{2M(1+\alpha)}{r}+\frac{M^2\alpha(1+\alpha)}{r^2} \,.
\label{eq:metricmodifiedgravity}
\end{eqnarray}
Here, $\alpha$ is a parameter governing the strength of the effective gravitational coupling $G=G_N(1+\alpha)$, representing a universal hair.

In this context, the function $g(r)$ is described by:
\begin{equation} \label{eq:example_hay}
\alpha g(r)=\ln \left[\frac{-\alpha ^2 M^2-\alpha  M^2+2 \alpha  M r+2 M r-r^2}{r (2 M-r)}\right].
\end{equation}

The seed metric is assumed to be the Schwarzschild one, meaning $X(r)=2M$. By substituting $r_{ph}^{(0)}=3M$ into \eqref{eq:example_hay}, we obtain:

\begin{equation}
\alpha g\left (r_{ph}^{(0)}\right)=\ln\left[1+\frac{\alpha ^2}{3}-\frac{5 \alpha }{3}\right]<0,
\end{equation}

In other words, the MOG parameter is constrained to be of the order $\alpha \leq 0.01$ \cite{bib:vanozi}, resulting in an increase in the size of a black hole shadow, as demonstrated in \cite{bib:vanozi}.

\section{Conclusion}

In conclusion, our exploration into the understanding of shadow and photon spheres surrounding black holes, especially in light of recent advancements such as those observed through the Event Horizon Telescope (EHT), underscores the significance of accounting for accretion discs in astrophysical scenarios. In this study, we proposed a model integrating small deformations via extended gravitational decoupling, thereby incorporating the presence of an accretion disc. By focusing on the effects of these deformations on the metric component $g_{00}$ and subsequent light propagation, we have revealed insights into the behavior of shadows and photon spheres. 

\begin{itemize}
    \item Theorem 1: If the seed metric is the Schwarzschild one, then the additional matter field increases the radius of a photon sphere if $g'(r_{ph}^{(0)})>0$ and decreases it if $g'(r_{ph}^{(0)})<0$. 
    \item Theorem 2: We can observe that if $g\left(r_{ph}^{(0)}\right)>0$ , then the shadow size decreases due to the additional matter field. Conversely, if 
 $g\left(r_{ph}^{(0)}\right)<0$, then the shadow increases. 
\end{itemize}
Notably, our findings align closely with previous research considering static spherically-symmetric black holes with plasma. These results not only contribute to refining theoretical models of black hole shadows but also lay a foundation for extending our understanding to more complex scenarios, such as axi-symmetric black holes, through methodologies like extended gravitational decoupling.

\acknowledgements
V. Vertogradov thanks the Basis
Foundation (grant number 23-1-3-33-1) for the financial support. A. {\"O}. would like to acknowledge the contribution of the COST Action CA21106 - COSMIC WISPers in the Dark Universe: Theory, astrophysics and experiments (CosmicWISPers) and the COST Action CA22113 - Fundamental challenges in theoretical physics (THEORY-CHALLENGES). We also thank TUBITAK and SCOAP3 for their support.



\end{document}